\documentclass[12pt]{article}
\pdfoutput=1
\usepackage{latexsym, graphicx,cite,mathrsfs}
\usepackage{amsmath}
\usepackage{amssymb}
\usepackage{amsthm}
\usepackage{subcaption}

\usepackage[top = 1. in,bottom = 1 in, left = 1 in, right=1 in]{geometry}
\usepackage[colorlinks=true, linkcolor=blue, bookmarks=true]{hyperref}
\newcommand{\arXiv}[1]{\href{http://www.arXiv.org/abs/#1}{arXiv:#1}}
\newcommand{\doi}[2]{\href{http://dx.doi.org/#2}{#1}}

\makeatletter
\renewcommand\section{\@startsection {section}{1}{\z@}%
                  {-3.5ex \@plus -1ex \@minus -.2ex}%nn
                  {2.3ex \@plus.2ex}%
                  {\normalfont\large\bfseries}}
\renewcommand\subsection{\@startsection{subsection}{2}{\z@}%
                   {-3.25ex\@plus -1ex \@minus -.2ex}%
                   {1.5ex \@plus .2ex}%
                   {\normalfont\bfseries}}
\makeatother

\numberwithin{equation}{section}

\newcommand{\beq}{\begin{equation}}
\newcommand{\eeq}{\end{equation}}
\newcommand{\ber}{\begin{array}}
\newcommand{\eer}{\end{array}}
\newcommand{\del}{\partial}

\newcommand{\al}{\alpha}
\newcommand{\bea}{\begin{eqnarray}}
\newcommand{\eea}{\end{eqnarray}}

\newcommand{\nn}{\nonumber}
\newcommand{\floor}[1]{\lfloor #1 \rfloor}

\begin{document}
\begin{titlepage}
\begin{flushright}
\phantom{arXiv:yymm.nnnn}
\end{flushright}
\vspace{5mm}
\begin{center}
{\LARGE\bf A~Gaussian~integral~that~counts~regular~graphs}\\
\vskip 17mm
{\large Oleg Evnin$^{a,b}$ and Weerawit Horinouchi$^a$}
\vskip 7mm
{\em $^a$ High Energy Physics Research Unit, Department of Physics, Faculty of Science, Chulalongkorn University,
Bangkok, Thailand}
\vskip 3mm
{\em $^b$ Theoretische Natuurkunde, Vrije Universiteit Brussel (VUB) and\\
International Solvay Institutes, Brussels, Belgium}
\vskip 7mm
{\small\noindent {\tt oleg.evnin@gmail.com, wee.hori@gmail.com}}
\vskip 20mm
\end{center}
\begin{center}
{\bf ABSTRACT}\vspace{3mm}
\end{center}

In a recent article {\it J. Phys. Compl. 4 (2023) 035005}, Kawamoto evoked statistical physics methods for the problem of counting graphs with a prescribed degree sequence. This treatment involved truncating a particular Taylor expansion at the first two terms, which resulted in the Bender-Canfield estimate for the graph counts. This is surprisingly successful since the Bender-Canfield formula is asymptotically accurate for large graphs, while the series truncation does not {\it a priori} suggest a similar level of accuracy.

We upgrade the above treatment in three directions. First, we derive an exact formula for counting $d$-regular graphs in terms of a $d$-dimensional Gaussian integral. Second, we show how to convert this formula into an integral representation for the generating function of $d$-regular graph counts. Third, we perform explicit saddle point analysis for large graph sizes and identify the saddle point configurations responsible for graph count estimates. In these saddle point configurations, only two of the integration variables condense to significant values, while the remaining ones approach zero for large graphs. This provides an underlying picture that justifies Kawamoto's earlier findings.

\vfill

\end{titlepage}

\section{Introduction}

Regular graphs are graphs of constant vertex degree. Thus, in a $d$-regular graph, all vertices are adjacent to exactly $d$ edges. Such graphs have often emerged as simple models of random networks in physics-oriented settings. For a sampler of literature, see \cite{reg1,qu,reg2,cycle,DSPL,TBK}.

The most basic question that arises before any more refined statistical considerations can be undertaken is counting the number of $d$-regular graphs on $N$ vertices.
This question turns out to be rather nontrivial, and no closed-form expression exists, though a variety of accurate asymptotic estimates have been established,
see \cite{Wormald, Noy} for the combinatorial perspective. In the statistical physics parlance, the number of graphs can be seen as the partition function
of a sort of microcanonical ensemble, while its logarithm is the corresponding `entropy.' Entropies of various random networks have been treated in \cite{entr1,entr3,entr4}.

In a recent article \cite{Kawamoto}, Kawamoto advocated applying techniques typical of evaluating statistical physics partition functions, like the Hubbard-Stratonovich transformation, to the problem of counting graphs with a prescribed degree sequence (an explicit list of vertex degrees), of which regular graphs are a special case. In \cite{Kawamoto}, one introduces a kind of `Lagrange multipliers' that enforce the prescribed vertex degrees, then the summation over all graphs can be performed, and one is left with an integral over these Lagrange multipliers. This is followed by applying an expansion of the sort  $\log(1+x)=x-x^2/2+\cdots$, truncated at the quadratic term, whereupon a Hubbard-Stratonovich transformation is evoked, and the result is expressed through Hermite polynomials. This result is shown to be compatible with the Bender-Canfield formula \cite{BC}, which is known to be asymptotically exact at large $N$. This degree of success is rather surprising, since the neglected terms in the expansion of the logarithm are not visibly suppressed at large $N$. An apparent paradox thus emerges that the treatment of \cite{Kawamoto} is more successful than what one would naively expect from the approximations that have been put in.

This essay is largely motivated by the desire to expose the underlying mathematical structure that supports the findings of \cite{Kawamoto}. To this end,
we shall avoid any truncated Taylor expansions and perform all computations exactly, specializing to the case of $d$-regular graphs so as not to clutter the formulas.
In this manner, we shall arrive at a curious exact representation for the number of $d$-regular graphs on $N$ vertices in the form of a $d$-dimensional Gaussian integral with a polynomial insertion. As an aside, this formula can be used in a handy manner to produce nice generating functions for the number of regular graphs.
We will then proceed with the large $N$ analysis by  applying conventional saddle point techniques. It will turn out that the saddle point configuration is rather simple, with only 2 out of $d$ possible directions actively participating, and the remaining ones effectively `switched off' at large $N$. These two `active' directions will reproduce the results of \cite{Kawamoto} in terms of a truncated logarithm expansion, but now without any assumptions beyond simply being in the large $N$ regime.

\section{An integral representation for the number of $d$-regular graphs}

As usual in analytic approaches to graph theory, we shall encode the graph connectivities of a graph on $N$ vertices in a symmetric, zero-diagonal, $N\times N$ {\it adjacency matrix} $\mathbf{A}$
whose components $A_{ij}$ equal 1 if there is an edge connecting vertices $i$ and $j$, and 0 otherwise. The $d$-regularity condition is then expressed simply as
\beq\label{defdreg}
\forall i:\quad \sum_{j=1}^N A_{ij}=d.
\eeq
To count all regular graphs, we sum over all possible adjacency matrices a function that equals 1 if (\ref{defdreg}) is satisfied, and 0 otherwise. To construct such a function, recall the evident complex contour integration identity:
\beq
\frac1{2\pi i} \oint\frac{dz}{z^{n+1}}=    \begin{cases}
       1&\mbox{if} \hspace{2mm} n=0, \\
       0&\mbox{otherwise},
    \end{cases}
\eeq
where the $z$-contour encircles the origin.
Then, we can write for the number of $d$-regular graphs on $N$ vertices:
\begin{equation} \label{Nz} 
\mathcal{N}_{N,d}=\sum_{\mathbf{A}}\oint\prod_{k=1}^N\left(\frac{dz_{k}}{2\pi iz_k^{d+1}}\,z_{k}^{\sum_{l=1}^NA_{kl}}\right).
\end{equation}
(Note that, in this counting, the vertices are labelled\footnote{Counting unlabelled graphs is a separate and rather intricate problem. The standard approach is via the P\'olya enumeration theorem, see chapter 4 of \cite{HP}. In this way, the counting is reduced to a sum over all partitions of the unlabelled vertex set. For a recent discussion from the standpoint of statistical physics, see \cite{unlabeled}.} in the sense that two graphs are counted as distinct if their adjacency matrices are distinct, even if the two matrices can be related to each other by row and column renumbering.)

The summation over $\mathbf{A}$ in (\ref{Nz}) means that we should sum over the values $0$ and $1$ each independent component of $\mathbf{A}$, that is, all $A_{kl}$ with $k<l$. To perform this summation explicitly, we first use the symmetry of $\mathbf{A}$ to re-express (\ref{Nz}) through such independent components as 
\begin{equation}
    \mathcal{N}_{N,d}=\sum_{\mathbf{A}}\oint\prod_{n}\frac{dz_{n}}{2\pi iz_n^{d+1}}\prod_{k<l}(z_{k}z_{l})^{A_{kl}}.
\end{equation}
Then, the summation over $\mathbf{A}$ is performed straightforwardly to yield
\beq\label{Nln}
    \mathcal{N}_{N,d}=\oint\prod_{n}\frac{dz_{n}}{2\pi iz_n^{d+1}}\prod_{k<l}(1+z_{k}z_{l})
    =\oint\prod_{n}\frac{dz_{n}}{2\pi iz_n^{d+1}}\exp\left[\sum_{k<l}\log(1+z_{k}z_{l})\right] .
\eeq
Such representations in terms of symmetric polynomials can often be noticed in combinatorial literature, see \cite{Mishna} for instance, and also \cite{MKW} in more immediate relation to regular graph counting. To a physicist's eye, the last exponential in (\ref{Nln}) looks like a sort of Boltzmann factor of a vector model defined in terms of $z_n$ with a logarithmic potential. Since the logarithmic potential is rather unwieldy for analytic manipulations, it is tempting to expand it as
\beq
    \log(1+z_{k}z_{l})=-\sum_{J=1}^{\infty}\frac{(-1)^{J}(z_{k}z_{l})^{J}}{J}. \label{Noint2}  
\eeq
Up to this point, we have been retracing the steps taken in \cite{Kawamoto}, but from now on our paths will diverge. In \cite{Kawamoto}, the expansion (\ref{Noint2}) was truncated to the first two terms, which made it possible to carry through the evaluation of (\ref{Nln}) in terms of Hermite polynomials, with results compatible with the Bender-Canfield formula -- a surprising level of accuracy given that the accuracy of truncating (\ref{Noint2}) to the first two terms was not protected by any small parameter. We, on the one hand, would like to avoid making approximations in (\ref{Noint2}), and are, on the other hand, facing the fact that handling the infinite expansion in (\ref{Noint2}) analytically would present a challenge. There is an important simplification that comes in, however: the truncation of (\ref{Noint2}) to the first $d$ terms, as opposed to the first two terms, is exact! This is because only positive powers of $z_k$ appear in (\ref{Noint2}), while any power of $z_k$ higher than $d$ inserted in the integral (\ref{Nln}) automatically makes it vanish. As a result, without making any approximations, we can set the upper bound of the $J$-summation in (\ref{Noint2}) to $d$. Furthermore, it is convenient to add and subtract the diagonal ($k=l$) terms to the $(k,l)$-sum in (\ref{Noint2}) to rewrite the expression equivalently as
\beq 
    \mathcal{N}_{N,d}=  \oint\prod_{n}\frac{dz_{n}}{2\pi iz_n^{d+1}} \exp\left[-\sum_{J=1}^{d}\frac{(-1)^{J}(\sum_{k}z_{k}^{J})^{2}}{2J}+\sum_{J=1}^{\floor{d/2}}\frac{(-1)^{J}\sum_{k}z^{2J}_{k}}{2J}\right]. \label{Noint3}
\eeq
At this point, a Hubbard-Stratonovich transformation comes in very handy: we introduce a set of auxiliary variables $\alpha_J$ with $J=1...d$ to write 
\beq
    \exp\left[-\frac{(-1)^{J}(\sum_{k}z_{k}^{J})^{2}}{2J}\right]=\frac{1}{\sqrt{2\pi J}} \int_{-\infty}^{\infty}\!\!\! d\alpha_{J}\, e^{-\alpha_{J}^2/2J}\exp\left[\frac{i^{J+1}\alpha_J\sum_{k}z_{k}^J}{J}\right].
\eeq
Once this is plugged back into \eqref{Noint3}, the integrals over $z_n$ completely factorize and can be evaluated independent of each other. To express the answer compactly, we define
\begin{equation}\label{Pdef}
    P_{d}(\al_{J})\equiv\frac{1}{2\pi i}\oint\frac{dz}{z^{d+1}}\exp\left(\sum_{J=1}^{d}\frac{i^{J+1}\alpha_{J} z^J}{J}+\sum_{J=1}^{\floor{d/2}}\frac{(-1)^{J}z^{2J}}{2J}\right).
\end{equation}
For any concrete $d$, this integral is evaluated straightforwardly in terms of residues expressed through the order $d$ $z$-derivative of the exponential, yielding an explicit polynomial\footnote{The polynomials defined by (\ref{Pdef}) can be related to cycle index polynomials of the symmetric group $S_d$, and further expressed through the complete exponential Bell polynomials \cite{Comtet}. We shall rely, however, on their explicit definition given in (\ref{Pdef}).} in terms of $\al_J$. This polynomial consists of monomial terms of the form $\prod_{J=1}^d \al_J^{p_J}$ with the powers involved satisfying $\sum_{J=1}^d Jp_J\le d$. With this polynomial at hand, we obtain the following exact representation for the regular graph count:
\beq\label{Nint}
\begin{split}
    \mathcal{N}_{N,d}&=\frac{1}{\sqrt{(2\pi)^d d!}} \int d\alpha_{1}\cdots d\alpha_{d}  \,e^{-\sum_{J=1}^{d}\alpha_{J}^2/2J}\left[P_{d}(\al_{J})\right]^N \\
   &= \frac{1}{\sqrt{(2\pi)^d d!}}\int d\alpha_{1}\cdots d\alpha_{d} \exp\left[-\sum_{J=1}^{d}\frac{\alpha_{J}^2}{2J}+N\log P_{d}(\al_J)\right].
\end{split}
\eeq
All of our subsequent studies will be based on this expression. What can be said about it? At concrete values of $N$, (\ref{Nint}) gives a nice way to actually compute the number of graphs. Namely, given $d$, $P_d$ is straightforwardly constructed, then raised to the power $N$, and after that, the Gaussian integral over $\al_J$ is taken without much effort for each individual monomial in $(P_d)^N$, using
\beq\label{eval}
\frac1{\sqrt{2\pi J}}\int_{-\infty}^\infty \!\!\!d\al_J\,\al_J^{2p} e^{-\al_J^2/2J}=J^p (2p-1)!!
\eeq
term-by-term.

We shall be mainly concerned below with two aspects of (\ref{Nint}). First, the simplicity of the dependence on $N$ in (\ref{Nint}) makes it attractive to convert this expression into a generating function for  $  \mathcal{N}_{N,d}$. This will be the subject of the next section. Thereafter, we will look for asymptotic estimates at large $N$. The second line of (\ref{Nint}) reveals an explicit structure of a saddle point integral, which suggests a natural way to analyze this formula at large $N$.

\section{Generating functions}

In constructing generating functions for regular graphs, a key concern is that the number of graphs grows very rapidly with $N$. Thus, the radius of convergence of the power series for naive generating functions will be zero, which, in application to (\ref{Nint}), will be reflected in divergences of the integral representation, making the formula largely useless.

For example, one could try an ordinary exponential-type generating function
\beq\label{Rexp}
\sum_N \frac{x^N}{N!}\mathcal{N}_{N,d}=\frac{1}{\sqrt{(2\pi)^d d!}} \int_{-\infty}^\infty\!\!\! d\alpha_{1}\cdots d\alpha_{d} \prod_{J=1}^d e^{-\alpha_{J}^2/2J}\exp[xP_{d}(\al_{J})].
\eeq
but this creates a divergent integral at $d>2$ since $P_d$ is a polynomial of degree $d$. If $P_d$ had only real coefficients, a practical remedy would have been to write
\beq
\sum_N \frac{(ix)^N}{N!}\mathcal{N}_{N,d}=\frac{1}{\sqrt{(2\pi)^d d!}} \int_{-\infty}^\infty\!\!\! d\alpha_{1}\cdots d\alpha_{d} \prod_{J=1}^d e^{-\alpha_{J}^2/2J}\exp[ixP_{d}(\al_{J})],
\eeq
but since $P_d$ has some imaginary coefficients, this fails for the same reason as the first attempt above.

We need to damp the behavior of graph counts at large $N$ stronger, which would correspondingly mitigate the behavior of $P_d$ at large $\alpha$. A possible approach is as follows:
\begin{align}\label{Rd}
R_d(x)&\equiv\sum_N \frac{x^N}{(dN)!}\mathcal{N}_{N,d}=\sum_N \frac{(\sqrt[d]{x})^{dN}}{(dN)!}\mathcal{N}_{N,d}\\
&=\frac{1}{\sqrt{(2\pi)^d d!}} \int_{-\infty}^\infty \!\!\!d\alpha_{1}\cdots d\alpha_{d} \prod_{J=1}^d e^{-\alpha_{J}^2/2J}\,\frac1{d}\sum_{q=0}^{d-1}\exp\left\{e^{2\pi i q/d}\left[xP_{d}(\al_{J})\right]^{1/d}\right\}.\nn
\end{align}
In this form, the large $\al$ behavior of  $P_d$ is tempered and the integral is convergent. The sum over $q$ can be seen as a generalization of the ordinary cosh function (which is recovered at $d=2$). Considerations of such `generalized hyperbolic' functions \cite{genhyp} apparently go back to the Riccati family. Note that, while the fractional powers involved in writing (\ref{Rd}) suffer from the usual ambiguities, these ambiguities completely drop out after the evaluation of the $q$-sum. The representation (\ref{Rd}) is rather suitable for numerical evaluation, for example, by running Monte-Carlo sampling based on the Gaussian measure.

In fact, it would suffice to use a smaller power instead of $d$. For example, we could use $d-1$ to obtain
\begin{align}\label{Rtd}
\tilde R_d(x)&\equiv\sum_N \frac{x^N}{((d-1)N)!}\mathcal{N}_{N,d}\\
&=\frac{1}{\sqrt{(2\pi)^d d!}} \int_{-\infty}^\infty\!\!\! d\alpha_{1}\cdots d\alpha_{d} \prod_{J=1}^d e^{-\alpha_{J}^2/2J}\frac1{d-1}\sum_{q=0}^{d-2}\exp\left\{e^{2\pi i q/(d-1)}\left[xP_{d}(\al_{J})\right]^{1/(d-1)}\right\}.\nn
\end{align}
The integral is still convergent.
This form has the advantage of connecting to the known exponential generating function at $d=2$ available in the literature, which also coincides at $d=2$ with (\ref{Rexp}). Indeed, 2-regular graphs have an extremely 
simple structure, being collections of disjoint loops of various sizes. This allows for a direct evaluation of their numbers by elementary methods. In terms of our current representation, one readily derives from (\ref{Pdef}) that
\beq
P_2(\al_1,\al_2)=\frac{\al_1^2}2-\frac{i\al_2+1}2.
\eeq
Then, from (\ref{Rtd}),
\beq
\tilde R_2(x)=\frac{1}{2\pi\sqrt{2}} \int_{-\infty}^\infty \!\!\!d\alpha_{1}\int_{-\infty}^\infty\!\!\! d\alpha_{2}\,e^{-\alpha_1^2/2-\alpha_2^2/4}\,\exp\left[x\left(\frac{\al_1^2}2-\frac{i\al_2+1}2\right)\right]=\frac{e^{-x/2-x^2/4}}{\sqrt{1-x}},
\eeq
which agrees with the result quoted in \cite{Noy}.

\section{A warm-up at $d=3$}\label{sec3reg}

In attempts to derive the large $N$ asymptotics of the second line in (\ref{Nint}) one encounters a subtlety. Naively, one could try to identify $\log P_d$ as the saddle point function and look for a saddle point estimate determined by its maxima. The problem is that, with respect to some $\alpha_J$ variables,
$P_d$ has no maxima at all. Most straightforwardly, $P_d$ is always linear in $\al_d$. In such situations, the saddle point must be found from the balance of $\log P_d$ and $\sum_J \al_J^2/2J$ terms in (\ref{Nint}), and the saddle point configuration will necessarily be $N$-dependent. Understanding the structure of this $N$-dependent saddle point will be our main goal in the remainder of this treatment. We shall start in this section with an explicit analysis of the simplest case $d=3$ that will help us identify the relevant patterns.

For 3-regular graphs, we obtain from (\ref{Pdef})
\beq\label{P3}
    P_{3}(\al_1,\al_2,\al_3)=\frac{1}{6} \left(-\alpha_{1}^{3}+3 \alpha_1\left(i \alpha_2+1\right)+2 \alpha_3\right).
\eeq
As a check, using this polynomial in (\ref{Nint}), we obtain by applying (\ref{eval}) the following counts for the number of 3-regular graphs on $N=4,6,8,10$ vertices, respectively:
1, 70, 19355, 11180820, in agreement with the numbers seen in \cite{Noy,OEIS}. The approach appears computationally rather effective, since evaluation at $N=100$ completes in just a few seconds on an ordinary personal computer using an analytic implementation for the Gaussian integrals in (\ref{Nint}).

Coming back to the problem of saddle point estimation of (\ref{Nint}), we must identify the saddle point configuration of $(\al_1,\al_2,\al_3)$ satisfying
\beq
    \frac{\partial}{\partial\alpha_{i}} \left(-\sum_{J=1}^{3}\frac{1}{2J}\alpha_{J}^2+N\log P_{3}\right) = 0.
\eeq
Explicitly, these are:
\begin{eqnarray}
   \alpha _1&=&\frac{ 3N(\alpha _1^2-i \alpha _2-1)}{\alpha_{1}^{3}-3 \alpha_1\left(i \alpha_2+1\right)-2 \alpha_3}, \label{a1} \\
    \frac{\alpha _2}{2}&=&-\frac{3 i N\alpha _1}{\alpha_{1}^{3}-3 \alpha_1\left(i \alpha_2+1\right)-2 \alpha_3},\label{a2} \\
    \frac{\alpha _3}{3}&=&-\frac{2 N}{\alpha_{1}^{3}-3 \alpha_1\left(i \alpha_2+1\right)-2 \alpha_3} .\label{a3}
\end{eqnarray}
Taking the ratios, we obtain
\begin{eqnarray}
    \frac{\alpha_{2}}{\alpha_{3}}&=&i\alpha_{1},\\
    \frac{2\alpha_{1}}{\alpha_{3}}&=&-\alpha _1^2+i \alpha _2+1 =-\alpha _1^2-\alpha _1\al_3+1.\label{db2} 
\end{eqnarray}

It is crucial for us to determine the leading power scalings of the $\al$'s in terms of $N$. Assume first that the r.h.s.\ of (\ref{db2}) is dominated by $+1$. In that case, $\al_3\sim\al_1$, $\al_2\sim\al_1^2$. Then, from (\ref{a2}), $\al_2\sim\sqrt{N}$, $\al_1^2\sim\sqrt{N}$, which contradicts the assumption that $+1$ dominates the r.h.s.\ of (\ref{db2}). Assume then that $\al_1\al_3$ dominates the r.h.s.\ of (\ref{db2}). Then, $\al_3\sim 1$, $\al_2\sim\al_1$, and from (\ref{a2}), $\al_1\sim\sqrt{N}$, so that the $\al_1^2$ term on the r.h.s.\ of (\ref{db2}) dominates the $\al_1\al_3$ term, in contradiction with our assumption. Assume finally that the $\al_1^2$ term dominates the r.h.s.\ of (\ref{db2}). In that case, $\al_3\sim 1/\al_1$, $\al_2\sim 1$, and from (\ref{a2}), $\al_1\sim\sqrt{N}$, so that the first term on the r.h.s.\ of (\ref{db2}) indeed dominates. Thus, only the solution
\beq
\al_1\sim\sqrt{N},\qquad \al_2\sim 1,\qquad \al_3\sim 1/\sqrt{N}
\eeq
is consistent.
We therefore write $\al_1=a_1\sqrt{N}$, $\al_2=a_2$ and $\al_3=a_3/\sqrt{N}$ and plug that back in (\ref{a1}-\ref{a3}) keeping only the leading terms as $N$ goes to infinity. This yields
\beq
   a _1^2=3,\qquad 
    a _2=-{6 i }/{a_1^2},\qquad
    a _3=-{6}/{a_{1}^{3}}.
\eeq
Thus, there are two saddle point configurations for (\ref{Nint}):
\beq\label{twosddl}
\al^{\scriptscriptstyle\mathrm{(saddle)}}_1=\pm\sqrt{3N},\qquad \al^{\scriptscriptstyle\mathrm{(saddle)}}_2=-2i,\qquad \al^{\scriptscriptstyle\mathrm{(saddle)}}_3=\mp2/\sqrt{3N}.
\eeq
At large $N$, we can treat the saddle point value of $\al_3$ as 0. We first focus on the saddle point with $\al_1=+\sqrt{3N}$ and expand in terms of `fluctuations' as
\beq
\al_1=\sqrt{3N}+\tilde\al_1,\qquad \al_2=-2i+\tilde\al_2,\qquad \al_3=\tilde\al_3.
\eeq
Then, for the saddle point function in (\ref{Nint}), neglecting any terms suppressed at large $N$, we get:
\beq
- \frac{\al_1^2}2-\frac{\al_2^2}{4}-\frac{\al_3^2}{6}+N\log P_3(\al_1,\al_2,\al_3)
=-\frac{3N}2-2+N\log(-N\sqrt{3N}/2)-\tilde\al_1^2-\frac{\tilde\al_2^2}{4}-\frac{\tilde\al_3^2}{6},
\eeq
The Gaussian integral over $\tilde\al_1$,  $\tilde\al_2$ and $\tilde\al_3$ is then evaluated as
\beq
\frac1{\sqrt{(2\pi)^33!}}\int_{-\infty}^\infty\!\!\! d\tilde\al_1d\tilde\al_2d\tilde\al_3\,e^{-\tilde\al_1^2-{\tilde\al_2^2}/{4}-{\tilde\al_3^2}/{6}}=\frac1{\sqrt{2}}.
\eeq
Repeating the same procedure for the second saddle point in (\ref{twosddl}) and adding up the results, we get
the following saddle point estimate of (\ref{Nint}), up to $1/N$ corrections:
\beq\label{sddl3}
\mathcal{N}^{\scriptscriptstyle\mathrm{(saddle)}}_{N,3}=\frac{e^{-2}}{\sqrt{2}}\left[\left(\frac{N\sqrt{3N}}{2e^{3/2}}\right)^{\!\!N}+\left(-\frac{N\sqrt{3N}}{2e^{3/2}}\right)^{\!\!N}\,\right].
\eeq
When $N$ is odd, this gives zero as it should, since there are no 3-regular graphs on odd numbers of vertices.
When $N$ is even, (\ref{sddl3}) agrees with the ``pairing model'' estimate seen in \cite{Noy}, which is identical to the Bender-Canfield formula applied to the case of regular graphs:
\beq\label{BCform}
\mathcal{N}^{\scriptscriptstyle\mathrm{(BC)}}_{N,d}=\frac{(dN-1)!!}{(d!)^N}e^{-(d^2-1)/4}=\frac{e^{-(d^2-1)/4}(dN)!}{2^{dN/2}(dN/2)!(d!)^N}\approx \sqrt{2}e^{-(d^2-1)/4}\left(\frac{(dN)^d}{e^d(d!)^2}\right)^{N/2}.
\eeq
where on should substitute $d=3$. The Bender-Canfield formula is known to be asymptotically exact at large $N$ (in the sense that the ratio of the exact graph number and this estimate approaches one), and so is our saddle point estimate (\ref{sddl3}). Indeed, we can compute the exact number of graphs with, say, N=200, as outlined under (\ref{P3}), and compare it with (\ref{sddl3}). Both computations give numbers around $6.4\times 10^{564}$, while the difference between the two results is around $1\%$.

\section{The large $N$ limit at general $d$}

For general $d$, $P_d$ should be recovered from (\ref{Pdef}) via a residue computation that gives
\beq\label{Pdiff}
    P_{d}(\al_1,\ldots,\al_d)=\frac{1}{d!}\frac{d^d}{dz^d}\exp\Bigg(\sum_{J=1}^d\frac{i^{J+1}\alpha_Jz^J}{J}+\sum_{J=1}^{\floor{d/2}}\frac{(-1)^{J}z^{2J}}{2J}\Bigg)\Bigg|_{z=0} .
\eeq
What can be said about the structure of this polynomial? If we differentiate the term involving $\al_J$ inside the exponential with respect to $z$, it brings down a factor of $z^{J-1}$ outside the exponential. Then, $J-1$ further $z$-derivatives will have to act on this factor to prevent the result from vanishing at $z=0$. This means that the powers $p_J$ inside any monomial $\prod \al_J^{p_J}$ in $P_d$ must satisfy $\sum_j Jp_J\le d$, as already mentioned under (\ref{Pdef}). The equality is attained if no $z$-derivatives act on the $\al$-independent terms inside the exponential in (\ref{Pdiff}).

We then turn to the saddle point equations for (\ref{Nint}) in the form
\beq\label{sddld}
\frac{\alpha_{k}}{k}=N\,\frac{\del_{\alpha_{k}}P_d}{P_d}. 
\eeq
Taking the ratios, as in the previous section, yields
\beq\label{ratiod}
    \frac{\alpha_{k}}{k\,\alpha_{1}}=\frac{\del_{\alpha_{k}}P_d}{\del_{\alpha_{1}}P_d}.
\eeq
Our first concern is, again, to identify the leading scalings of the $\al$'s within the solution at large $N$. While in this general case, it would be burdensome to run the complete balance analysis assuming that one or another term in the sums dominates, we can use the considerations of the previous section as a guidance, propose that all sums are dominated by terms with the highest power of $\al_1$, and check the consistency of the resulting solution. The terms with the highest powers of $\al_1$ in $P_d$ can be symbolically extracted as follows (omitting the coefficients):
\begin{eqnarray}
    P^{(\max \al_1\!)}_{d} &\sim& \alpha_{1}^d+\alpha_{1}^{d-2}\alpha_{2}+\alpha_{1}^{d-3}\alpha_{3}+\cdots+\al_1\al_{d-1}+\alpha_{d}.
\end{eqnarray}
Retaining only these terms, (\ref{ratiod}) reduces to $\al_k\sim \al_1^{2-k}$, while (\ref{sddld}) with $k=1$ yields $\al_1\sim\sqrt{N}$. We thus arrive at the saddle point scalings
\beq\label{sddlNd}
\al_1\sim \sqrt{N},\qquad \al_2\sim 1,\qquad \al_k\sim \frac1{N^{k/2-1}}.
\eeq
This pattern is consistent with our starting assumption that the highest powers of $\al_1$ dominate all sums in the numerator and denominator of (\ref{sddld}). Indeed, replacing any factors of $\al_1$ with higher $\al$'s obviously lowers the power of $N$. Note that inserting $\al_J$ in a monomial in $P_d$ requires removing at least $J$ powers of $\al_1$ to comply with the constraints on power counting stated under (\ref{Pdiff}).

At large $N$, $\al_k$ with $k>2$ can be {\it effectively set to zero} in the saddle point configuration (\ref{sddlNd}). The large $N$ limit is thus governed by $\al_1$ and $\al_2$ alone.
To determine their detailed form, we need the coefficients of $\al_1^d$ and $\al_1^{d-2}\al_2$ in $P_d$. These are straightforwardly extracted to yield,
also including the $\al_1^{d-2}$ term that we shall need below:
\beq\label{Ptrunc}
P_d=\frac{(-\al_1)^{d-2}}{d!}\left(\al^2_1-\frac{d(d-1)}2(i\al_2+1)\right)+\cdots.
\eeq
With this, we obtain two saddle point configurations:
\beq\label{sddlgen}
\al^{\scriptscriptstyle\mathrm{(saddle)}}_1=\pm\sqrt{dN},\qquad \al^{\scriptscriptstyle\mathrm{(saddle)}}_2=-(d-1)i,\qquad \al^{\scriptscriptstyle\mathrm{(saddle)}}_{k\ge 3}=0.
\eeq
As in the previous section, we first focus on the $+\sqrt{dN}$ saddle point and introduce fluctuations as
\beq
\al_1=\sqrt{dN}+\tilde\al_1,\qquad \al_2=-(d-1)i+\tilde\al_2,\qquad \al_{k\ge 3}=\tilde\al_{k}.
\eeq
Then, keeping only terms nonvanishing at large $N$, for which it suffices to truncate $P_d$ to the terms given in (\ref{Ptrunc}), we obtain
\beq
- \sum_{J=1}^d\frac{\al_J^2}{2J}+N\log P_d(\al_1,\ldots,\al_d)\\
=-\frac{dN}2-\frac{d^2-1}4+N\log\frac{(-\sqrt{dN})^{d}}{d!}-\tilde\al_1^2- \sum_{J=2}^d\frac{\tilde\al_J^2}{2J}.
\eeq
For the remaining Gaussian integral over $\tilde\al$, we get
\beq
\frac{1}{\sqrt{(2\pi)^d d!}} \int d\tilde\alpha_{1}\cdots d\tilde\alpha_{d}  \,e^{-\tilde\al_1^2+\sum_{J=2}^{d}\tilde\alpha_{J}^2/2J}=\frac1{\sqrt{2}}.
\eeq
Combining this with the contribution from the second saddle point in (\ref{sddlgen}), we obtain the following estimate of (\ref{Nint}):
\beq\label{Nsddld}
\mathcal{N}^{\scriptscriptstyle\mathrm{(saddle)}}_{N,d}=\frac{e^{-(d^2-1)/4}}{\sqrt{2}}\left[\Bigg(\frac{(\sqrt{dN})^d}{e^{d/2}d!}\Bigg)^{\!\!N}+\Bigg(\frac{(-\sqrt{dN})^d}{e^{d/2}d!}\Bigg)^{\!\!N}\,\right].
\eeq
This, again, agrees with the prediction of the ``pairing model'' \cite{Noy} and the Bender-Canfield formula (\ref{BCform}), which are known to be asymptotically accurate. In our treatment, an asymptotic expansion is generated by construction through our application of the saddle point method to (\ref{Nint}), and the result can be further improved, if necessary, with $1/N$ corrections.

The fact that the saddle point configurations (\ref{sddlgen}) has vanishing $\al_{k\ge 3}$ provides an underlying picture for the surprising accuracy
of the derivations in \cite{Kawamoto}. There, the expansion (\ref{Noint2}) was truncated at $J=2$. We did not do this truncation, but rather treated the expansion exactly, introducing the auxiliary variables $\al_J$ for the different terms of the $J$-expansion to facilitate the $z$-integrations. As a result,
a saddle point problem for $\al_J$ emerged, and its solution revealed vanishing values for $\al_J$ with $J\ge 3$ at $N\to\infty$. Thus, our asymptotic result (\ref{Nsddld}) does not, in fact, receive contributions from the terms in (\ref{Noint2}) with $J>2$. While this structure is evident in our implementation of the saddle point method, it is not obviously seen at the level of the expansion (\ref{Noint2}) as used in \cite{Kawamoto}, since higher order terms are not suppressed manifestly by powers of $N$. It is only after the saddle point problem in terms of $\al_J$ has been set up that the underlying structure becomes apparent.

\section{Conclusions}

We have developed a family of multivariate polynomials (\ref{Pdef}), in terms of which the numbers of labelled $d$-regular graphs on $N$ vertices are exactly expressed by the Gaussian integrals (\ref{Nint}). These integrals can be converted to attractive generating functions (\ref{Rd}) and (\ref{Rtd}) for regular graph counts, expressed through integrals of generalized hyperbolic functions and Gaussians. We have finally implemented a saddle point treatment of the integral (\ref{Nint}) at large $N$, obtaining an accurate asymptotic estimate for the number of graphs (\ref{Nsddld}) compatible with the ``pairing model'' and the Bender-Canfield formula. As usual in saddle point settings, the result can be systematically improved to incorporate $1/N$ corrections, though computations will become progressively more involved at higher subleading orders.

The use of Gaussian integrals, Wick contractions and Feynman diagrams to count discrete geometric structures has a venerable history,
in particular, in relation to counting two-dimensional triangulations \cite{Ambjoern,GM}. Most broadly, our Gaussian representation for regular graph counting can be seen as part of that trend \cite{kazakov, BJK,higherd, SasakuraSato}.

In more practical terms, the `auxiliary field' representation we have developed, coupled with the subsequent saddle point evaluation at large $N$ is
very much in line with the methods of `statistical field theory' \cite{statFT,statFTneu} and `large $N$' analysis \cite{largeN} that have been previously applied 
to a wide range of problems involving random matrices and random graphs, see for instance \cite
{BR,FM,MF,euclRM,PN,kuehn,spectrum,AC,resdist,RP2022,laplace}. Because of the extensive number of constraints on vertex degrees,
random regular graphs are not straightforward to accommodate in this framework, and yet our current treatment spells out a procedure that 
can be used to handle them successfully. This, in turn, creates various prospects for using statistical field theory methods to treat diverse
spectral and random walk problems on random regular graphs, and more.\vspace{5mm}

\section*{Acknowledgments}

We thank Tatsuro Kawamoto, Pawat Akara-pipattana and especially Sergei Nechaev for valuable discussions about regular graph counting and related topics.
 OE is supported by Thailand NSRF via PMU-B, grant number B13F670063. 
WH is supported by the NSTDA of Thailand under the JSTP scholarship, grant number SCA-CO-2565-16992-TH.\vspace{1cm}

%%%%%%%%%%%%%%%%%%%%%%%

\end{document}